\PassOptionsToPackage{expansion=false}{microtype}

\documentclass[]{vrk}

\usepackage[T1]{fontenc}
\usepackage{lmodern}
\usepackage{textcomp}

\usepackage{hyperref}
\usepackage{url}
\usepackage[most]{tcolorbox}
\usepackage{xcolor}
\usepackage{colortbl}
\usepackage{booktabs}
\usepackage{listings}
\usepackage{fancyvrb}
\usepackage{siunitx}
\usepackage{subcaption}
\usepackage{caption}
\usepackage{xcolor}
\usepackage{listings}
\usepackage[most]{tcolorbox}
\tcbuselibrary{listings,skins,breakable}
\usepackage[most]{tcolorbox}
\tcbuselibrary{listings,skins,breakable}
\usepackage{graphicx}
\usepackage{tabularx}
\usepackage{array}
\usepackage{pgf-pie}

\usepackage[T1]{fontenc}
\usepackage{lmodern}
\usepackage{textcomp}

\usepackage{hyperref}
\usepackage{url}
\usepackage{xcolor}
\usepackage{colortbl}
\usepackage{booktabs}
\usepackage{listings}
\usepackage{fancyvrb}
\usepackage{siunitx}
\usepackage{subcaption}
\usepackage{caption}
\usepackage{enumitem}
\usepackage[most]{tcolorbox}
\usepackage{ragged2e}
\tcbuselibrary{listings,skins,breakable}
\usepackage{graphicx}
\usepackage{tabularx}
\usepackage{array}

\title{
Design Conductor 2.0: An agent builds a TurboQuant inference accelerator in 80 hours
}

\author{The Verkor Team: Ravi Krishna, Suresh Krishna, David Chin}

\abstract{Driven by a rapid co-evolution of both harness and underlying models, LLM agents are improving at a dizzying pace. In our prior work (performed in Dec. 2025), we introduced ``Design Conductor'' (or just ``Conductor''), a system capable of building a 5-stage Linux-capable RISC-V CPU in 12 hours \cite{team2026design}. In this work, we introduce an updated multi-agent harness powered by frontier models released in April 2026, which is able to handle 80x larger tasks, at higher quality, fully autonomously. Following a brief introduction, we examine 4 designs that the system produced autonomously, including ``VerTQ'', an LLM inference accelerator which hard-wires support for TurboQuant in a 240-cycle pipeline, starting from the TurboQuant arXiv paper \cite{zandieh2025turboquant}. VerTQ includes heavy compute processing, with 5,129 FP16/32 units; the design was mapped to an FPGA at 125 MHz and consumes \SI{5.7}{\milli\metre\squared} in TSMC 16FF (8 attention pipes). We review the key new characteristics that enabled these results. Finally, we analyze Design Conductor's token usage and other empirical characteristics, including its limitations.}

\date{\today}
\correspondence{\email{team@verkor.io}}

\begin{document}

\maketitle

\begin{figure*}[h!]
	\centering
	\includegraphics[width=0.85\textwidth]{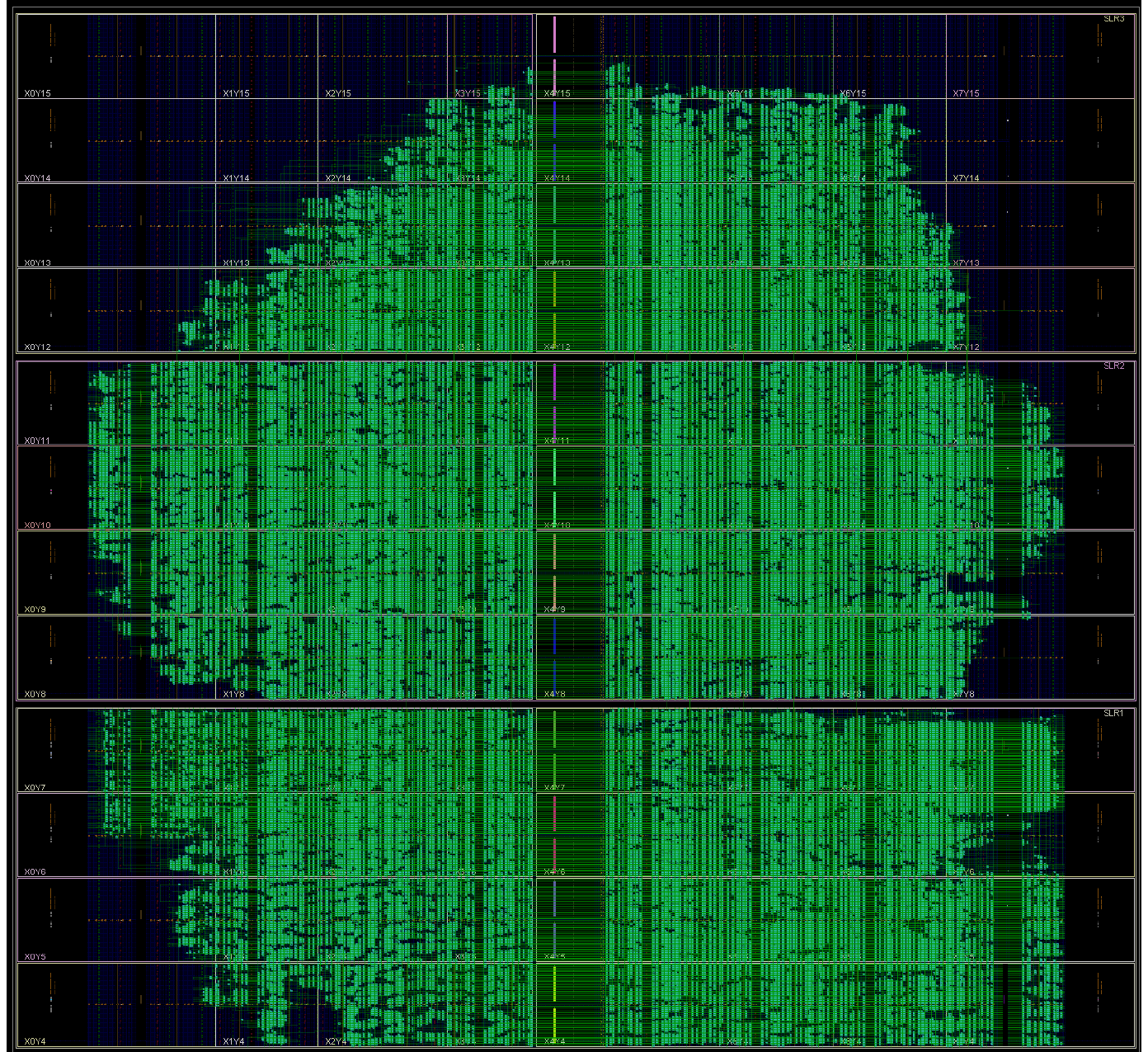}
	\caption{\textbf{VerTQ Physical Layout in 4-die XCVU29P-3 FPGA. 3x SLR dies shown. Conductor 2.0 optimized the architecture to minimize inter-die signal crossings. }}
	\label{fig:vertq_fpga}
\end{figure*}

\section{Introduction}
\label{sec:intro}

Building a new chip and taking it to ``tape-out'', or the manufacturing stage, is today an extremely expensive and slow process. It involves many steps, starting from the architecture, RTL and testbench implementation at the frontend, verification, synthesis, place \& route, power optimization, and packaging and thermals analysis. All of these steps in combination result in a process that costs over \$400M and consumes 18-36 month for teams of hundreds of people (who typically start with an existing design). Any issues identified during the process can require additional design iterations, and in the worst case, ``re-spins'' of the silicon, requiring new masks.

Long-horizon autonomous AI agents present the promise of changing this paradigm. However chips, unlike software, are physical products, and this necessitates a different agent design than is optimal for coding. For one, the chip design process contains many steps spanning both logical and physical aspects, and all of these must be handled for an agent to be useful and avoid being bottlenecked. Because chips are manufactured at great upfront cost (with an N2 mask set estimated at >\$30M), functional verification takes on an even greater importance than in software. Bugs cannot be fixed by pushing a patch, and so instead painstaking efforts must be expended to ensure all cases are covered from the outset.

In our prior work \cite{team2026design}, we introduced ``Design Conductor'', a long-running multi-agent system designed to solve these problems. In that work, it was able to build a (comparatively simple) 5-stage RISC-V CPU. The agent ran for around 12 hours and used the RISC-V ISA simulator, Spike \cite{spike} as a reference model for system sim. We highlighted several limitations of Design Conductor in that work, including its architecture reasoning, understanding of timing, and need for very precise specification. In all of these regards, the Design Conductor of that time (approximately Dec. 2025) was more like a highly skilled and inexhaustible implementer than a true designer. Some in the industry also observed that openly available implementations of RISC-V cores may have been included in the pretraining data of the frontier models that Conductor uses, making the task ``easier'' for the agent to do\footnote{In the authors' opinion, this is a specious point, but we acknowledge it nonetheless.}.

\begin{figure*}[!htbp]
	\centering
	\includegraphics[width=\linewidth, trim=0.4cm 0.5cm 0cm 3.5cm, clip]{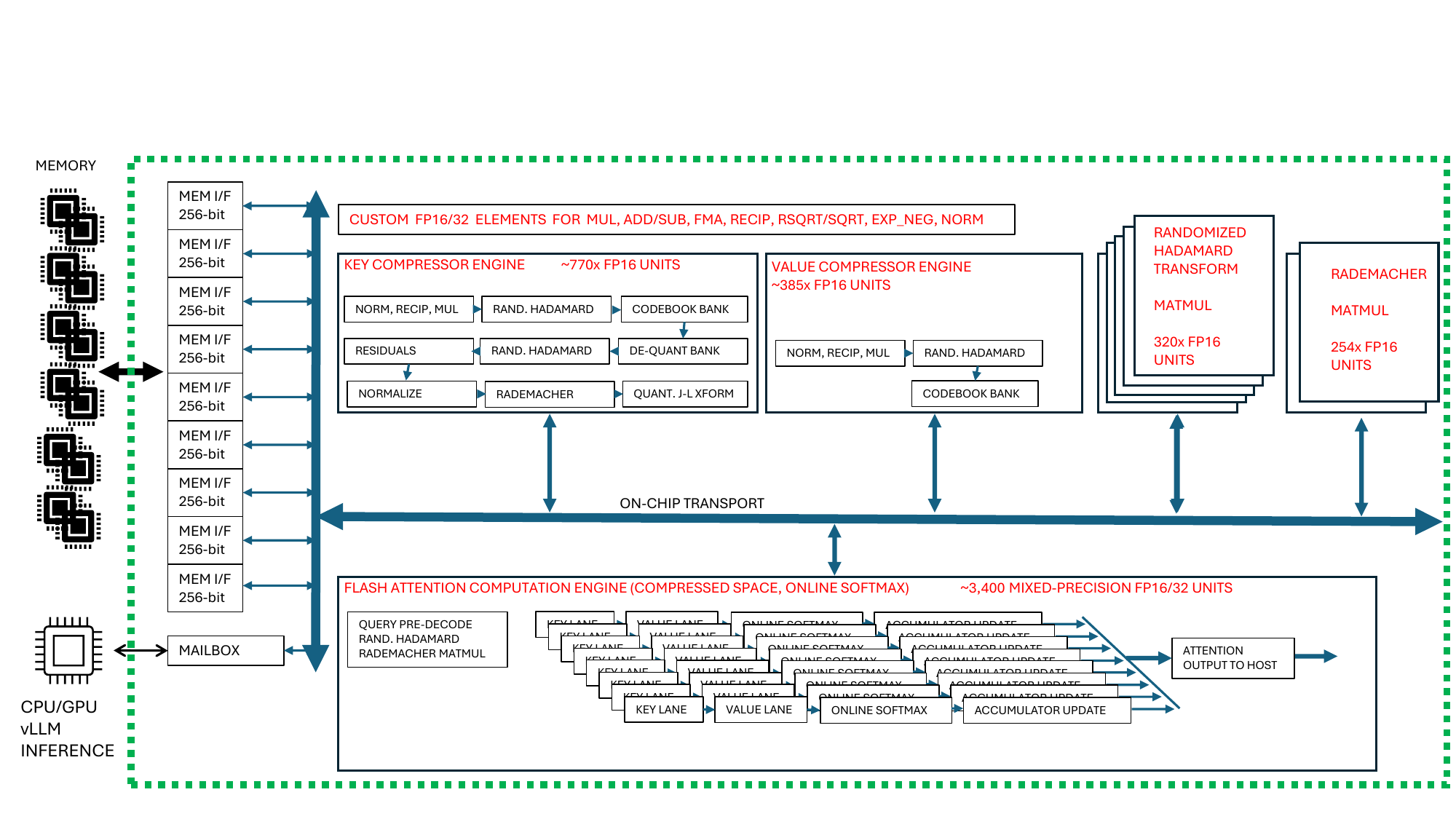}
	\caption{\textbf{``VerTQ'' TurboQuant/FlashAttention Chip by Conductor 2.0 - Block Diagram}}
	\label{fig:vertq_block}
\end{figure*}

By redesigning the Conductor harness and taking advantage of new frontier models released in April 2026, we find that the system has been able to make major inroads in each of these areas, and others. We tested this ``Design Conductor 2.0'' on many design tasks as part of our evaluation process, and we present a subset of 4 of these designs in this work. The most interesting such task was assigning Design Conductor to build ``VerTQ'', an inference accelerator for TurboQuant \cite{zandieh2025turboquant}, a KV-cache compression algorithm. It effectively architected, implemented, verified, timing-optimized, and mapped to an FPGA an accelerator for the entire LLM decode operation (both the QK and V matmuls and the online softmax) which hard-wires support for TurboQuant. To our knowledge, there is no such hardware available online (or anywhere). Figure \ref{fig:vertq_block} shows a block diagram of VerTQ.

In building VerTQ, Design Conductor demonstrated architecture judgment and the ability to guide and manage a complex project over a roughly 80-hour runtime. We believe this capability -- ``concept to layout'' -- will enable major changes in the fabless semiconductor ecosystem.
\section{DC-built Designs}
\label{sec:designs}

This section presents the key characteristics and methodology for four designs generated by Conductor 2.0.

\subsection{``VerTQ'' TurboQuant/FlashAttention Chip}

\begin{table}[h]
	\centering
	\small
	\begin{tabularx}{\linewidth}{
			@{}
			>{\raggedright\arraybackslash\hsize=0.5\hsize}X
			>{\raggedright\arraybackslash\hsize=1.5\hsize}X
			@{}
		}
		\hline
		\multicolumn{2}{c}{\textbf{VerTQ Key Features}} \\
		\hline \\
		Major Functions &
		K compression via TurboQuant-Prod with QJL residuals compression, V compression via TurboQuant-MSE, Decoder with Integrated FlashAttention pipeline \\
		
		Software Support &
		Direct integration into Python vLLM Inference framework \\
		
		Key Features &
		4.3x KV cache compression, 16x fewer multiplies inner attention loop, 9-bank memory interface \\
		Arithmetic & 5,129 mixed-precision FP16/FP32 arithmetic units (8-way attention decoder) \\ \\
		\hline \\
		Chip Size (FPGA), single attention decoder &
		500,619 LUT, 247,022 FF, 748 DSP48E2, 4 RAMB36, 9 RAMB18 \\
		
		Chip Size (FPGA), 8-way attention decoder &
		$\sim$1.9M LUTs, $\sim$300K FF, $\sim$1.5K DSP48E2, 18 RAMB36, 9 RAMB18 \\
		
		Chip Size (16nmFF est.) &
		$\sim$2 mm$^2$ single attention decoder, $\sim$5.7 mm$^2$ 8-way, $\sim$18 mm$^2$ 32-way \\
		
		Operating Frequency &
		125 MHz (XCVU29P-3 FPGA)\\ \\
		\hline
	\end{tabularx}
	\caption{VerTQ Key Features}
	\label{tab:vertq_features}
\end{table}

\subsubsection{Key Features}

VerTQ is an accelerator chip that implements Google's TurboQuant algorithm \cite{turboquant_google_blog} which reduces memory usage of Large Language Models by a factor of 4.3x while maintaining full performance and quality. This algorithm directly enables LLM inference applications to use less memory (which is currently in short supply) and to run faster as it saves precious memory bandwidth. Google announced the TurboQuant algorithm (in mathematical form) on March 24, 2026; as of publication date there is not, to the best of our knowledge, any hardware implementation of TQ. Google's TQ announcement had a direct effect on the stock prices of leading memory chip manufacturers \cite{kharpal2026google_turboquant_memory_stocks}.

VerTQ ``sits'' between a main processor and the DRAM memory system, effectively managing the KV-cache on behalf of the processor. VerTQ handles the compression of KV data and, additionally, accelerates the computationally burdensome attention computation by performing FlashAttention \cite{dao2022flashattention}  operations on-chip, including online softmax. These operations are performed without decompressing KV-cache data, which further saves memory bandwidth. VerTQ is specifically intended for edge AI applications, such as automobiles, drones, and robots, where compact size, power/compute efficiency and low-cost are key.

Figure \ref{fig:vertq_fpga} shows the layout of VerTQ implemented in a Xilinx XCVU29P-3 FPGA. Figure \ref{fig:vertq_block} shows a block diagram of VerTQ. This device consists of 4 dies, each known as a ``Super Logic Region'' (SLR). Notice that VerTQ fits into 3 SLRs and that inter-SLRs crossings, which involve slow signal crossings between dies, are minimized. This figure illustrates how Conductor 2.0 considered technology limitations up-front, during the architecture/spec phase, in order to produce a chip that maps optimally to available technology resources.

Table \ref{tab:vertq_features} shows the key features of VerTQ. This design is intended to be suitable for ``edge AI'' accelerator applications that require autonomous, on-device LLM support. Owing to its relatively small die area and high compute efficiency, VerTQ is suitable for integration into such an accelerator. The integration process can be handled entirely by Conductor 2.0, given a few words of user input describing the requirements.

\subsubsection{Input to Conductor 2.0}

The following user input was given to Conductor 2.0, along with the TurboQuant references and an existing software implementation. Based on this input, Conductor 2.0 autonomously performed the following tasks over an 80-hour period:

\begin{itemize}
	\item Architecture exploration
	\item Python vLLM Inference Framework integration
	\item Module-level design and micro-architecture
	\item Physical design including synthesis and P\&R
	\item Module-level verification
	\item System simulation including integration with vLLM\\
\end{itemize}

\begin{tcolorbox}[
	colback=white,
	colframe=black,
	coltext=vrkblue,
	breakable,
	fontupper=\scriptsize\fontfamily{pcr}\selectfont,
	before upper={\RaggedRight}
	]
	\textbf{VerTQ Chip Design -- User Input}
	
	\smallskip
	
	We want to build a hardware implementation of TurboQuant in the form of an internal block named VerTQ. The ultimate target is an FPGA, Xilinx XCVU29P-3. Vivado is available.
	
	\smallskip
	
	Consider how we should structure VerTQ and set its interfaces -- the current thinking is as a block in front of memory which writes/reads KV cache entries, indices, or pages and handles the TQ decompression/compression as it does so, and also handles attention score computation for a single query used during autoregressive decoding.
	
	\smallskip
	
	We want this hooked up to an inference engine such as vLLM. We'd basically plug this into where the typical KV cache is managed in those inference frameworks. For now, we would do this with simulation, with actual FPGA mapping and connection to be added later. You'll need to functionally verify VerTQ using the software implementation of TurboQuant.
	
	\smallskip
	
	While SW simulation won't stress throughput, we want the engine to be capable of operating at a high throughput in actual use. In this scenario, the engine has to be able to write compressed results and read decompressed results as fast as you can make the hardware run given the constraints of the XCVU29P-3 FPGA.
	
	\smallskip
	
	Target a 125 MHz internal clock rate for VerTQ, using Vivado out of context and the most aggressive Vivado synthesis and place-and-route timing options.
	
	\smallskip
	
	Use available on-chip DSP blocks if appropriate to do floating-point arithmetic. If you build your own floating point arithmetic, simplify the units by implementing DAZ and FTZ. Also, we don't care about generating exceptions/errors for invalid floating point data or underflow/overflow. In general, we don't care about handling any error conditions, so don't implement any error handling logic.
	
	\smallskip
	
	A couple notes:
	\begin{itemize}
		\item We want to use 3 PQ bits, plus 1-bit QJL for K.
		\item We want to demo with Qwen3-4B.
	\end{itemize}
	
	Let's use FP16 for the KV cache precision. You'll need to confirm this works with vLLM.
	
	\smallskip
	
	The host that runs vLLM will be simulated as well; it will write values and read results from a dedicated part of the simulated SRAM, and operations will be started/stopped by the simulation harness. Based on the interface that you have defined to vLLM, the simulation harness will handle moving data back and forth and coordinating hardware simulation with running vLLM.
	
	\smallskip
	
	Best of luck building VerTQ!
\end{tcolorbox}

\subsubsection{Verification}

Conductor 2.0 wrote extensive testbenches for each module, using reference Python implementations to debug functionality and cycle-by-cycle behavior. For system sim, Conductor used data captured from Qwen3-4B. Due to limited simulation servers, Conductor could only run a test case involving all 36 layers of Qwen3-4B with a limited context size of 64. The availability of an emulation system or a server farm could make it possible to run larger samples e.g. context size of 32k.

\subsubsection{Challenges}

A lot of debug time, and the corresponding high volume of tokens, was consumed troubleshooting floating-point numerics. In order to optimize hardware implementation, Conductor 2.0 used FP16 operators where possible, and FP32 where needed (e.g. online softmax accumulation). As a result, there are minor deviations from a Python reference implementation of Qwen3-4B. The resulting discrepancies could not be ignored, as they may reflect a hardware or design bug. Each discrepancy had to be inspected, analyzed and traced back to its root cause, which is most cases turns out to be reasonable accuracy/cost trade-offs. Some situations, such as input vectors with high norm, may introduce larger discrepancies. Conductor's decision to build a custom library of highly optimized floating point elements, meant to optimize timing beyond what the FPGA's DSPs could do and tuned to the specific number ranges found in LLM applications, compounded the potential for issues with numerics. An example is a high-speed specialized exponentiation unit that Conductor built, specifically optimized for the distribution of values found in the online softmax computation of FlashAttention. Conductor did find a design bug in this unit, whereby a polynomial used to implement negative exponentiation turned out to have excessive error in some situations. Conductor replaced this with a higher degree, fifth-order Taylor polynomial evaluated via Horner's algorithm.

Timing closure was a significant challenge as well on this design, due to the wide datapaths that carry significant state through long pipelines. The resulting data routing is a challenge in any technology, but particularly so in an FPGA.

\subsection{Heavily pipelined AES Core}

\subsubsection{Key Features}

Conductor 2.0 designed a very high throughput core for AES encryption. AES is an NIST-standard algorithm \cite{aes} that is widely used in the industry for encrypting data in transport and at rest. Conductor's AES core focuses on datacenter applications, hence there is a need to maximize performance. A typical AES encryption engine within a DPU or networking interface would consist of multiple such cores. Conductor's AES core operates in CTR mode, with a 128-bit key and achieves >100Gbps at 1GHz in a smaller KU5P-3 FPGA, while achieving >400Gbps at 3.2 GHz in an ASIC process (ASAP7 7nm PDK, TT process corner). These speeds compare favorably with commercially available AES single cores \cite{synopsys_aes_xts_ecb_ip}, \cite{altera_aes_128_ip}.

\subsubsection{Verification}

Conductor verified the core extensively using the NIST-validated Python \texttt{cryptography} library as a reference, invoked as \texttt{Cipher(algorithms.AES(key), modes.ECB())}. Conductor chose to run an extensive set of vectors, including \texttt{NIST SP 800-38A} and \texttt{FIPS-197 KAT}, plus about 100k random vectors.

\subsubsection{Implementation}

\begin{figure*}[t]
	\centering
	
	\begin{subfigure}[t]{0.49\textwidth}
		\centering
		\includegraphics[width=\linewidth]{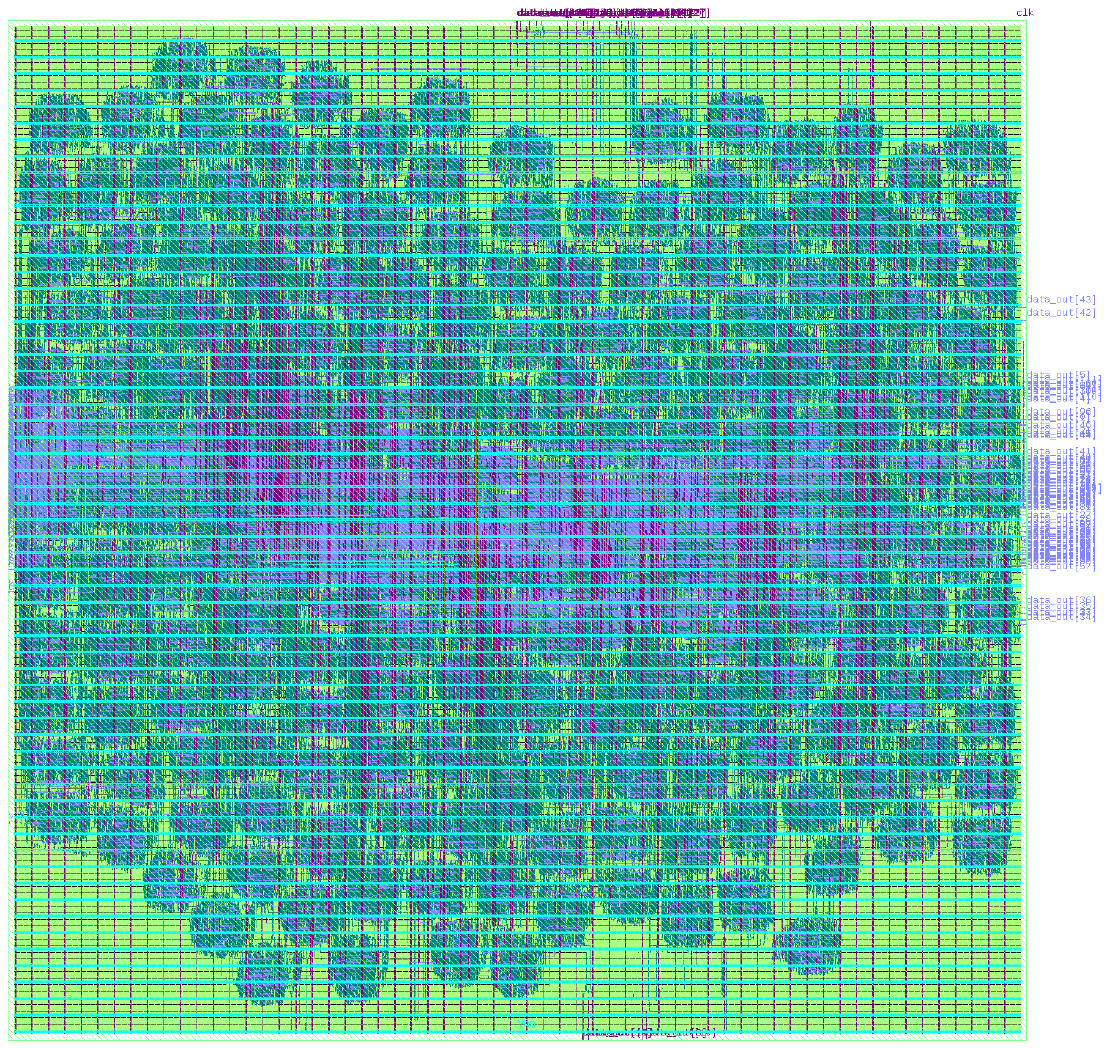}
		\caption{\textbf{Physical layout of AES core in ASAP7 7nm PDK. Note the visible 160 ``S-Boxes''. Die size is about \SI{333}{\micro\metre} on a side.}}
		\label{fig:aes_asap7}
	\end{subfigure}
	\hfill
	\begin{subfigure}[t]{0.49\textwidth}
		\centering
		\includegraphics[width=\linewidth]{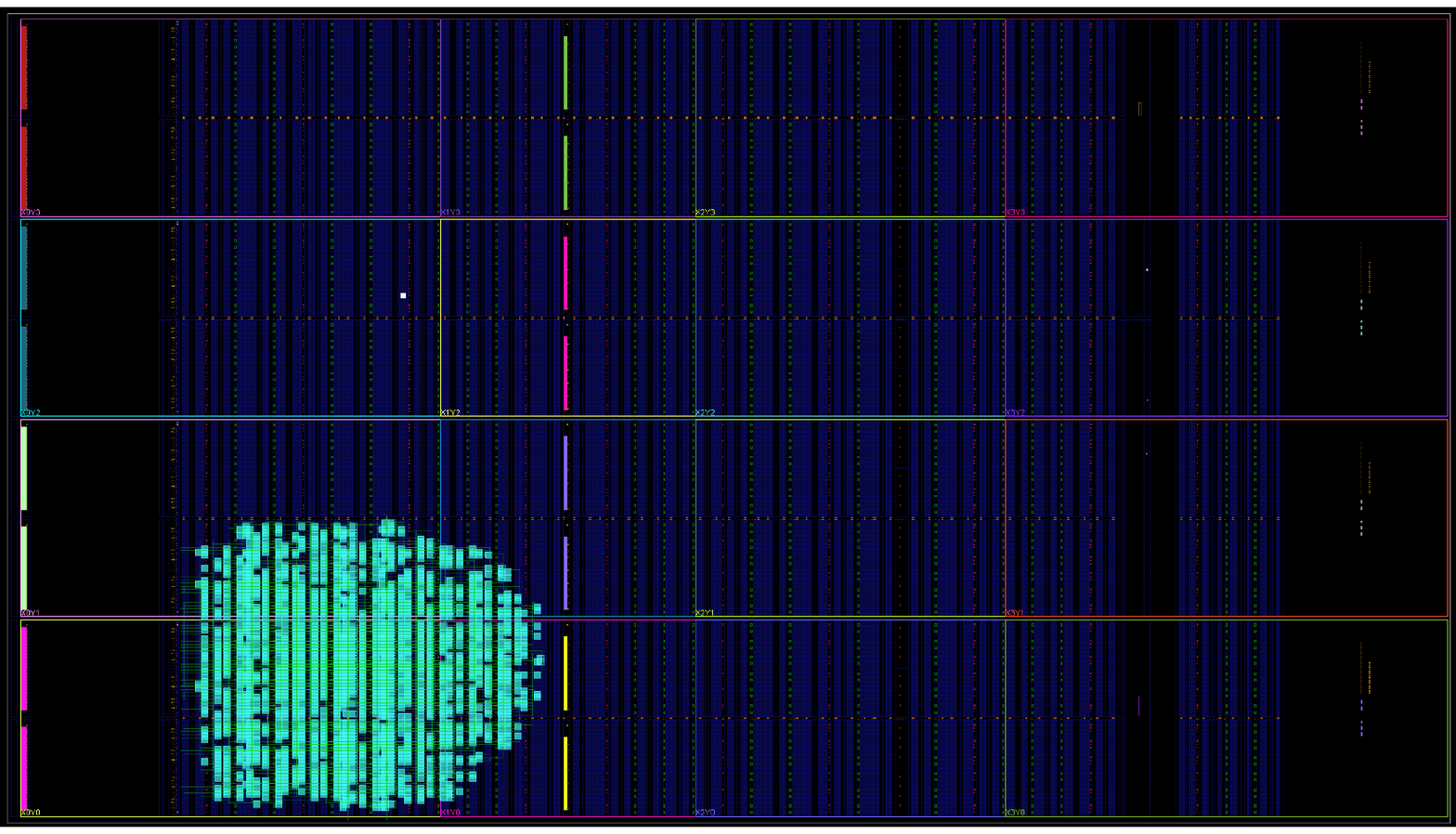}
		\caption{\textbf{Physical layout of AES core in Xilinx KU5P-3 FPGA.}}
		\label{fig:aes_fpga}
	\end{subfigure}
	
	\caption{\textbf{AES physical layouts in ASIC and FPGA implementations.}}
	\label{fig:aes_layouts}
\end{figure*}

Figure \ref{fig:aes_asap7} shows the GDSII for Conductor's AES core using the ASAP7 7nm PDK. The design is a spatially unrolled implementation for maximum performance, as it is pipelined to accept one 128-bit data word per clock cycle (after a brief key setup period). The 160 AES encryption ``S-Boxes'' are clearly visible in the layout. Figure \ref{fig:aes_fpga} shows the layout of the AES core in a Xilinx KU5P-3 FPGA.

It is interesting to note that Conductor chose significantly different designs for the FPGA vs. 7nm versions of the AES core. The ASAP7 7nm version implements a 60-deep pipeline that results from micro-pipelining the ``S-Boxes''. This implementation maximizes clock rate in the ASAP7 7nm PDK process, but maps poorly to an FPGA where gates are implemented via LUTs. For the FPGA version, Conductor chose to map the ``S-Boxes'' to FPGA LUTs, using 8x4 LUT6 per 256-entry ``S-Box''. This micro-architecture makes optimal use of the structure of the Xilinx FPGAs and results, for that technology, in a much higher clock rate and throughput.

For Conductor, refactoring RTL at will is a straightforward and reliable process. For human engineers, changing the structure of a deeply pipelined design would represent considerable, error-prone effort.

\subsubsection{Input to Conductor 2.0}

The following lists the user input that was given to Conductor 2.0:

\begin{tcolorbox}[
	colback=white,
	colframe=black,
	coltext=vrkblue,
	breakable,
	fontupper=\scriptsize\fontfamily{pcr}\selectfont,
	before upper={\RaggedRight}
	]
	\textbf{Heavily Pipelined AES core -- User Input}
	\par\smallskip
	
	here's the task, using [GITHUB REPO URL]:
	\par\smallskip
	
	I would like to implement a maximum throughput AES encryption engine in Verilog. Inputs are clock, reset, a 128-bit key, a single-cycle valid pulse, and a 128-bit data block that is driven when valid is high. Output is a 128-bit encrypted block and a valid signal for one clock. We only need CTR mode.
	\par\smallskip
	
	The encryptor must include the key schedule generation logic. The keying material is set once, and remains the same as multiple 128-bit blocks go through the encryptor pipeline.
	\par\smallskip
	
	Only throughput matters. Latency does not matter. Power and area do not matter.
	\par\smallskip
	
	Target the xcku5p-ffva676-3-e device. Vivado should be available.
	\par\smallskip
	
	Use OpenSSL as the golden reference.
	\par\smallskip
	
	\textbf{(Later on)} Now implement the same design using OpenROAD/YoSys, which should be available on your machine, with the ASAP7 7nm process library. Work off the same repo, but structure the RTL and synthesis directories to cleanly separate the new work, while sharing files as needed. You will need to modify the S-Box design; the LUT version will not achieve good clock rates on the ASAP7 library. You will need to pipeline the S-Boxes deeply using various techniques. You can re-use parts of your previous work, available here: [GITHUB REPO URL]
\end{tcolorbox}

\subsection{Optimized FP32 Add/Sub}

\subsubsection{Key Features}

In this design, Conductor 2.0 was asked to build a pipelined 32-bit floating point add/sub unit with a throughput of one result per clock cycle and a very aggressive clock rate goal in a Xilinx KU5P-3 FPGA. This design is interesting in that it illustrates Conductor's ability to relentlessly optimize a special function unit, taking into account the specifics of backend implementation technology, as well as every form of ``tips and tricks'' to optimize the FPGA implementation.

Conductor 2.0 was able to achieve a clock rate of 896 MHz on a Xilinx KU5P-3 FPGA, with an 11-stage pipeline that is capable of producing one result per clock cycle. The unit uses 437 LUTs, 586 FF and 12 CARRY8. In terms of clock rate and pipeline depth, Conductor's design compares favorably with commercially available FPGA-optimized FP32 add/sub IP from Xilinx such as their Floating-point v7.1 library \cite{amd_floating_point_v7_1_perf}. Conductor's design achieves a higher clock rate, similar pipeline depth, and similar resource utilization though slightly higher LUT count. Like the commercial component, Conductor's design implements DAZ and FTZ policies for handling floating-point denorms.

\subsubsection{Verification}

Verification represents the major challenge for any floating-point unit, including this one. Conductor devoted most of its effort (and tokens) to generating an extensive test suite, leveraging Python's \texttt{numpy.float32} validated library to code a bit-accurate model of the expected behavior of the fp32 add/sub unit. Conductor generated ~918k test vectors including random sets and corner-cases for validating floating-point units. The runtime of this design was fairly long relative to its very small size as shown in Table \ref{tab:relative-usage-runtime}, owing to the extensive verification required.

\subsubsection{Implementation}

Conductor performed essentially a guided ``grid search'' of available backend synthesis and P\&R options, generating a multitude of TCL scripts with functions to encapsulate various backend strategies. Conductor was able to learn from these experiments and select the best strategy, which was key to achieving a superhuman clock rate.

\begin{figure*}[h!]
	\centering
	\includegraphics[
	width=0.85\textwidth,
	trim=0 140bp 0 140bp,
	clip
	]{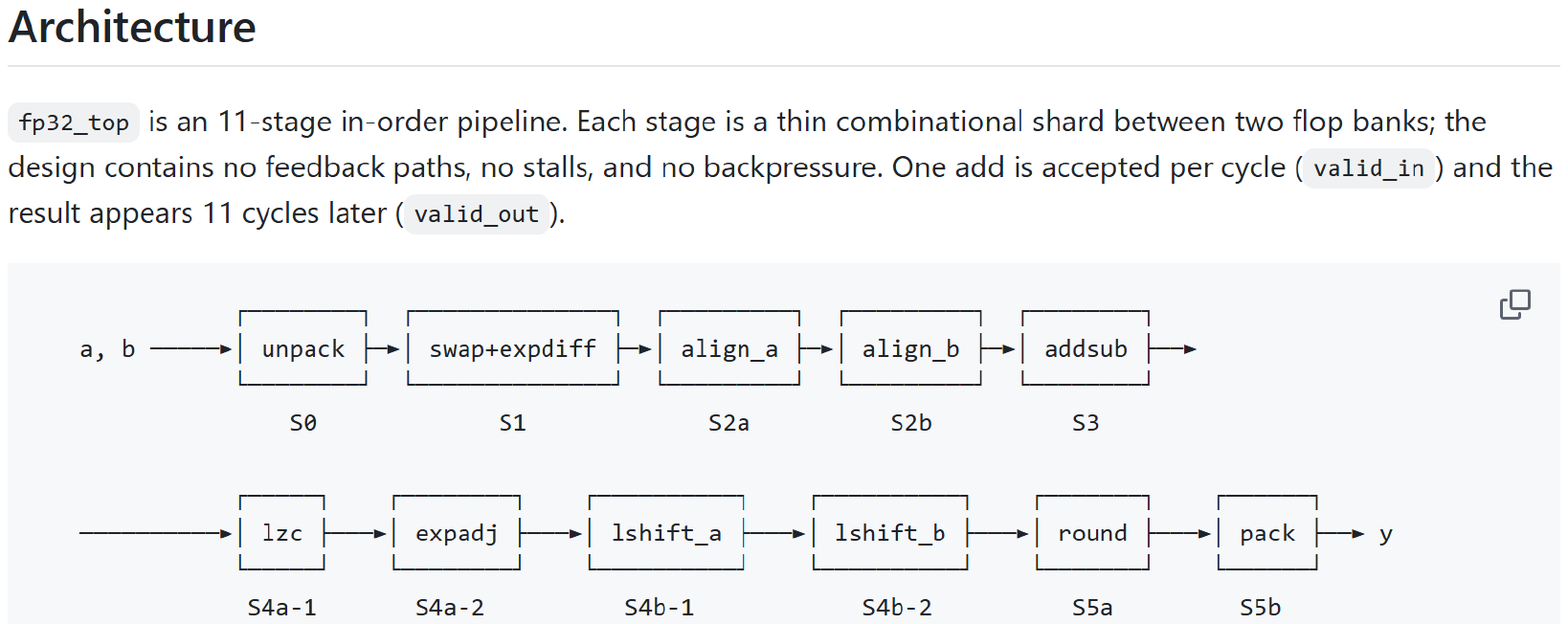}
	\caption{\textbf{Conductor's description of its fp32 add/sub pipeline.}}
	\label{fig:fp32pipe}
\end{figure*}

Conductor generated an ASCII block diagram of the pipeline structure that it converged on, as shown in Figure \ref{fig:fp32pipe}.

\subsubsection{Input to Conductor 2.0}

The following lists the user input that was given to Conductor 2.0 for this design task:

\begin{tcolorbox}[
	colback=white,
	colframe=black,
	coltext=vrkblue,
	breakable,
	fontupper={\scriptsize\fontfamily{pcr}\selectfont},
	before upper={\RaggedRight}
	]
	\textbf{Pipelined FP32 add/sub -- User Input}
	
	\smallskip
	
	The task is to build an FP32 adder targeting the XCVU19P. Vivado should be available. Target is 500 MHz. Use this repo: [GITHUB REPO URL].
	
	\smallskip
	
	RNE only is fine; denormal as zero is fine; 1 add/cycle but the latency/stage count is of your choice.
	
	\smallskip
	
	\textbf{(Later on)} Change the device to xcku5p-ffva676-3-e and change the Vivado options. Target 800 MHz clock rate.
\end{tcolorbox}

\subsection{Line rate allreduce added to switch core}

\subsubsection{Key Features}

The allreduce design differs from previous examples in that Conductor was given very detailed instructions to retrofit the allreduce feature into an existing VOQ switch core based on the Corundum open-source design \cite{corundum_ethernet_switch}. This design illustrates Conductor's ability to analyze a legacy design and ``bolt on'' major new features in an optimized manner, using existing data and control interfaces. The new feature, in-network allreduce, is a ``lite'' version of NVIDIA's SHARP \cite{graham2016scalable} and is a key component used for accelerating the distributed training of large neural networks such as LLMs.

The specifics of the design that Conductor completed are as follows:

\begin{itemize}
	\item 64-bit AXI-Stream data bus.
	\item 4 input ports × 4 output ports.
	\item Extensive arbitration \& control logic to handle FIFO buffering, stalls, packet input \& output delays.
	\item Allreduce: BF16 sum-reduction, with FP32 accumulation, via magic header 0xA11F3D00.
	\item Four allreduce aggregators, located near the output ports.
	\item Four mixed precision BF16/FP32 reduction accumulators per aggregator.
	\item Total 16x mixed-precision BF16/FP32 adders, 8-stage pipelined, handles normals/subnormals/specials.
	\item Double-buffered allreduce aggregators to support full line rate allreduce.
	\item Up to 512 BF16 elements per port per allreduce operation.\\
\end{itemize}

\subsubsection{Verification}

Conductor developed testbenches to verify functionality, correctness, to check corner conditions such as combinations of various packet types, various patterns of backpressure, and bubble insertions.
In addition, Conductor verified the new design's ability to sustain line-rate allreduce. In aggregate, Conductor ran about 23K packets in simulation.

\subsubsection{Implementation}

The final implementation targets a Xilinx KU5P-3 FPGA and uses 14,236 LUTs, 18,568 FFs, 32 RAMB36 and 16 RAMB18 (the RAMs are mainly used for FIFOs such as VOQ).
A large portion of the CLBs are used for the pipelined floating-point reduction accumulators, which are mixed precision BF16 with FP32 accumulation; there are 4 such accumulators per aggregator, times 4 aggregators.

\subsubsection{Input to Conductor 2.0}

The following lists the user input that was given to Conductor 2.0 for this design task; these instructions are considerably more detailed than typically used for greenfield designs:

\begin{tcolorbox}[
	colback=white,
	colframe=black,
	coltext=vrkblue,
	breakable,
	fontupper={\scriptsize\fontfamily{pcr}\selectfont},
	before upper={\RaggedRight}
	]
	\textbf{Line-rate bf16/fp32 in-network allreduce -- User Input}
	\par\smallskip
	
	You should work off the repo [GITHUB REPO URL]. Run the existing testbenches there and run the entire switch through Vivado for the Xilinx XCKU5P-3FFVA676E device to establish a baseline Fmax. Vivado is installed on the machine and available to you. For the testbenches, use Verilator, which is available on the machine.
	
	\par\smallskip
	Important notes:
	
	\begin{itemize}[leftmargin=*, topsep=2pt, itemsep=1pt]
		\item Always use the \texttt{switch.v} design; ignore the 3 others.
		\item For floating point arithmetic, do not handle denorms. Instead, treat denorms as zeros.
		\item We don't care much about overall latency or added die area.
	\end{itemize}
	
	\par\smallskip
	As a helpful tip, note that the exponent bit count in bf16 and fp32 is the same, so you can just zero-pad the mantissa to convert, and the other way you just truncate the mantissa bits.
	
	\par\smallskip
	Please document your progress as you go in this Slack channel.
	
	\par\smallskip
	\textbf{FUNCTIONAL TASK:}
	
	\par\smallskip
	Add support for in-switch all-reduce, sum only; don't worry about division. Inputs are bf16, accumulated into fp32 buffers, and output packets are also bf16. This would be used in an AI cluster for gradient allreduce. You'll need to extend the existing tests to test this allreduce case. The new tests should include interleaving allreduce traffic with regular traffic.
	
	\par\smallskip
	The goal is to implement the new functionality such that it runs at the same Fmax as the baseline switch. Use Vivado for this, and adjust pipelining and RTL as needed.
	
	\par\smallskip
	Description of the design modifications below.
	
	\par\smallskip
	\textbf{Allreduce input packets:}
	
	\par\smallskip
	Input packets use a little-endian ``magic header'' as follows; this ``magic header'' is located inside the regular packet payload:
	
	\begin{itemize}[leftmargin=*, topsep=2pt, itemsep=1pt]
		\item Magic field, 32 bits, \texttt{magic\_hdr}, equal to \texttt{0xA11F3D00}.
		\item Element count, 32 bits, field \texttt{elem\_count}, which is the number of bf16 values in the input packet, from 1 to 512 maximum bf16 values.
		\item Aggregator number, 32 bits, \texttt{aggr\_no}, which is one of 0, 1, 2, or 3.
		\item Payload, which is \texttt{elem\_count} bf16 values.
	\end{itemize}
	
	\par\smallskip
	The output packet, which is the result of allreduce accumulation, should use the same format but the bf16 values represent the summed input values across 4 packets per aggregator, as defined below.
	
	\par\smallskip
	\textbf{Aggregators:}
	
	\begin{itemize}[leftmargin=*, topsep=2pt, itemsep=1pt]
		\item The design includes 4 aggregators. Each aggregator is attached to one output port, after that output's \texttt{axis\_arb\_mux}, after \texttt{switch\_crossbar}. Each aggregator holds 512 fp32 values. Each aggregator receives an input packet destined for it, based on the \texttt{aggr\_no} field, and accumulates \texttt{elem\_count} bf16 incoming values into the corresponding fp32 aggregator values.
		
		\item Since the switch uses a 64-bit internal bus, each beat carries 4 bf16 values. You therefore need to provide 4 adders for each aggregator, each adder capable of accumulating one fp16 value into the corresponding fp32 aggregator value. Each of these adders will need to run at line rate; I would suggest pipelining them deeply, perhaps 5 stages or more.
		
		\item At reset, the fp32 values in the aggregators are undefined. Don't bother zeroing them.
		
		\item Each aggregator should have a 3-bit input packet counter that keeps track of how many input packets it has processed, from 0 to 4. This counter resets to 0 and increments every time the aggregator receives an input packet.
		
		\item When an aggregator receives its first packet, i.e. its input packet counter is 0 at the start of the packet, it should write the values from the packet, bf16 converted to fp32, without attempting to aggregate. This can be accomplished by adding a mux between the adder output and the input to the aggregator's fp32 buffer memory to drive the packet values, converted to fp32, instead of the adder output, or by adding a mux to the input of the adder tree to drive zeros when the first packet is received.
		
		\item When an aggregator has received 4 packets, it should send out its accumulated result packet, converted from fp32 to bf16, on its corresponding output port, then zero out its input packet counter.
		
		\item Once an aggregator is ready to send out data, the switch should finish sending out any packets that are already in the process of being sent out on output ports, and then hold off on any further output packets such that all 4 output ports are kept available, and then give priority to the aggregator output, which is replicated across all 4 ports.
		
		\item If multiple aggregators are ready to send out data, round-robin should be used to arbitrate among them.
		
		\item Once an aggregator has sent out its result packet, the old fp32 values can remain in the aggregator; don't bother resetting or zeroing them. The first new packet will overwrite those values anyway.
		
		\item While an aggregator is sending out its packet data, it should be able to receive another input packet and start storing that and set its packet counter to 1. This requires each aggregator to have double-buffered packet storage.
		
		\item In terms of datapath muxing, it should be possible to allow either buffer of each double-buffered aggregator to send data to output ports. When a particular aggregator is sending out data, the data should be broadcast across all 4 output ports. This can be done using \texttt{axis\_broadcast} from \texttt{lib/verilog-axis/rtl}, or equivalently by directly writing Verilog code.
		
		\item You may implement aggregators as a self-contained unit that sits inline with each output port and includes, internally, a bypass path to allow regular packets through. Alternatively, you may implement aggregators as separate logic hanging off to the side of output ports, with appropriate muxing.
		
		\item An input packet can be accumulated into either buffer, but once accumulation starts into a particular buffer, that same buffer will be used until 4 packets are received and the results are sent out.
		
		\item Once results are complete, meaning 4 packets received, the aggregator should switch to the other double-buffer to receive more input packets.
		
		\item Use only one set of fp32/fp16 adders per aggregator. The adder tree should be shared among the two double-buffers. It is never the case that input packets need to be simultaneously accumulated into both buffers of an aggregator.
	\end{itemize}
	
	\par\smallskip
	\textbf{Test cases:}
	
	\begin{itemize}[leftmargin=*, topsep=2pt, itemsep=1pt]
		\item Create test cases that involve multiple back-to-back all-reduce packets, with a goal of demonstrating line-rate aggregation capability.
	\end{itemize}
\end{tcolorbox}

\section{New Capabilities}
\label{sec:capabilities}

\subsection{Building Blocks}
\label{subsec:blocks}

Long-running multi-agent systems typically share many basic structural attributes. The particular mechanisms by which each of these is implemented is what enables a system to excel at its target use case.
\begin{enumerate}
	\item Context window management: dealing with LLM sessions, optimizing for prefix hit rates, operating around context limits
	\item Subagent session management and control: structuring the tasks to get high-quality results
	\item Memory: maintaining state across multiple LLM sessions and over the duration of a project that would otherwise take a human team months
	\item Knowledge: many systems incorporate specialized knowledge or skills through various mechanisms
\end{enumerate}

In Design Conductor 2.0, all of these attributes were rebuilt to support longer-horizon and more complex semiconductor design tasks. 

\subsection{Novel Capabilities}

Based on the building blocks described in \ref{subsec:blocks}, and paired with models released in April 2026, Design Conductor 2.0 acquired several new capabilities that made it possible to build the designs described in section \ref{sec:designs}.

\subsubsection{Concept to Layout}

In contrast to the original Design Conductor 1.0, the new version is able to develop its own specifications. It is able to take a concept -- like a TurboQuant accelerator -- and autonomously map it to a performant design. This means that it can actually be a true partner to master designers, and that the right interface is no longer spec to RTL or spec to layout, but rather concept onwards. Design Conductor 2.0 can relieve the team from much of the work of building exhaustive and authoritative specifications.

\subsubsection{Architecture understanding \& first-principles reasoning}

One of the primary areas where Design Conductor 1.0 struggled was in making reasoned architecture and system design tradeoffs. This constrained its applicability in more complex designs where engineers need to make these tradeoffs on a regular basis. The new harness effectively solves this problem.

For example, in the case of VerTQ, it developed a custom exponentiation unit that uses a (to our knowledge) novel factoring of a fifth-degree polynomial to minimize delay at the precision the task required. It considered how to structure and map the VerTQ design to the VU29P multi-die FPGA, minimizing communication overhead. This ability to reason from first principles is what makes it possible for Design Conductor 2.0 to tackle completely new problems with innovative solutions.

\subsubsection{Closing the loop}

Thanks to the other capabilities noted above, Design Conductor 2.0 has the ability to ``close the loop'', taking high-level feedback from timing analysis and placement tools, reasoning about it, and making complex RTL changes to adapt to this. In the case of VerTQ, we tested this by modifying the timing target once an initial version was built. When this happens, Conductor refactors large parts of the design to achieve the PPA goals. This blunts the impact of one of the most painful experiences for an all-human design team -- finding out that a design isn't going to meet timing and needs to be restructured.

\subsubsection{Mathematical Reasoning Abilities}

Conductor leverages the improved mathematical reasoning abilities of the new models. In particular, Conductor 2.0 is able to accurately analyze design characteristics, as well as to optimize the implementation of numerics for specific applications.

\section{System Characteristics \& Limitations}
\label{sec:results}

\subsection{Empirical Characteristics}

\begin{table}[t]
	\centering
	\begin{tabular}{lcc}
		\hline
		\textbf{Design} & \textbf{Relative token usage} & \textbf{Relative wall-clock runtime} \\
		\hline
		VerCore (Design Conductor 1.0) & 1.0x & 1.0x \\
		Optimized FP32 add/sub & 2.1x & 0.75x \\
		In-switch allreduce & 0.8x & 0.3x \\
		AES encryptor & 4x & 4x \\
		VerTQ & 12x & 6.7x \\
		\hline
	\end{tabular}
	\caption{Relative token usage and wall-clock runtime by design.}
	\label{tab:relative-usage-runtime}
\end{table}

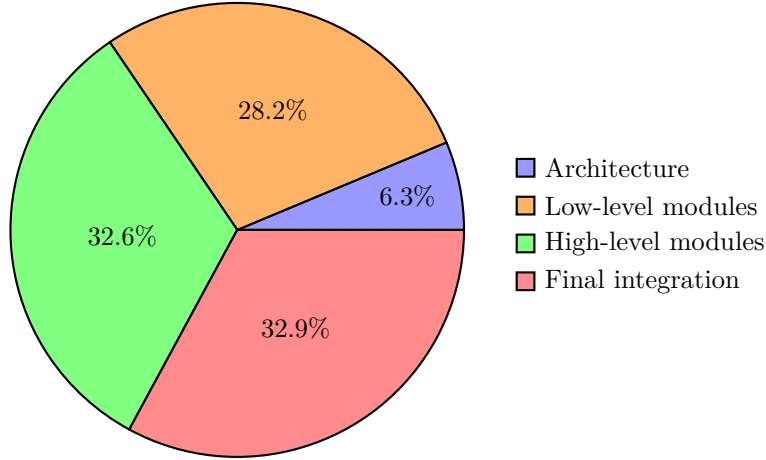
\begin{figure}
	\centering
	\begin{tikzpicture}
	\pie[
	text=legend,
	radius=3,
	color={blue!40, orange!60, green!50, red!45}
	]{
		6.3/Architecture,
		28.2/Low-level modules,
		32.6/High-level modules,
		32.9/Final integration
	}
	\end{tikzpicture}
	\caption{Token usage breakdown (dollar normalized)}
	\label{fig:tok_breakdown}
\end{figure}

See table \ref{tab:relative-usage-runtime} for relative token usage statistics (these are normalized to dollar token cost as comparing raw input, output, cached input, etc. is relatively meaningless).

Figure \ref{fig:tok_breakdown} shows the token usage breakdown. Not unlike a human team, verification and achieving timing closure are the most expensive components.

\subsection{Limitations}

\subsubsection{Overly Methodical}

We find that the system can be overly methodical or over-cautious when dealing with tasks that are actually as simple as they seem. This manifests when asking the system simple queries -- for example, how many floating-point units there are in the design. We observe cases where instead of responding quickly with the obvious answer it already knows, the system will do an extensive codebase review, checking and re-checking its work many times, only to come to the same answer after keeping the user waiting for 20 minutes.

We are unsure if this is due to the new frontier models the updated harness uses or if it's an inherent harness artifact. 

\subsubsection{Backend vs. uArch fixes}

We observed cases where Design Conductor would propose overly complex RTL changes to problems that could be resolved with simple backend script fixes.

\subsubsection{Goal-setting}

We found that in terms of overall constraint and goal-setting, Design Conductor was sometimes too aggressive -- for example, aiming for too high a clock frequency target. 

\subsubsection{Human Review Bottleneck}

While Design Conductor 2.0 is capable of handling very large end-to-end tasks, we find, as many teams on the agent frontier have, that human review is the major bottleneck. The only ``solution'' we have found to this is for the human designers to think through the key design constraints (PPA) upfront.

\clearpage
\newpage

\bibliographystyle{plain} 
\bibliography{ref}

\end{document}